\documentclass[12pt,amsmath,amssymb,]{revtex4}
\usepackage{graphics}
\usepackage{amsmath}
\usepackage{epsfig}
\usepackage{epsf}
\usepackage{graphicx}
\usepackage{dcolumn}
\usepackage{bm}

\begin{document}
\title{The third Zemach moment and the size of the proton}

\date{\today}

\author{Bea Ya Wu and Chung Wen Kao \\
Department of Physics, Chung-Yuan Christian University,
Chung-Li 32023, Taiwan \\ }

\begin{abstract}
To resolve the puzzle of the proton size raised from the recent result of muonic hydrogen Lamb shift, De R\'{u}jula has proposed
that a large value of the third Zemach moment $\langle r^3_{p}\rangle_{(2)}$ of the proton to be the solution.
His suggestion has been criticized by many groups based on the $ep$ scattering data at low $Q^2$ regime.
However, if there is a ``thorn'' or ``lump'' in the electric form factor of the proton $G_{E}(Q^2)$ at extremely low $Q^2$ regime, then
the third Zemach moment $\langle r^3_{p}\rangle_{(2)}$ would be as large as De R\'{u}jula suggested.
In this article, we show that the existence of such a ``thorn" or ``lump'' has not been completely excluded, although tightly restricted, by the current data of $ep$ elastic scattering. We also suggest a more sophisticated global fitting procedure of $G_{E}(Q^2)$ for the future fitting.
\end{abstract}

\maketitle

\section{Introduction}
The issue of the charge radius of the proton has attracted a lot of attention, since the charge radius extracted from
the Lamb shift of muonic hydrogen has been reported to be $0.84184(67)fm$ \cite{Pohl}.
This result is significantly smaller than the previous value of CODATA \cite{CODATA}
, $\sqrt{\langle r_{p}^2\rangle}$(CODATA)=$0.8768(69)$ fm and the one extracted from $ep$ elastic scattering data $\sqrt{\langle r_{p}^2\rangle}$(ep)=$0.879(5)_{stat}(4)_{syst}(2)_{model}(4)_{group}$ fm \cite{A1}.
De R\'{u}jula \cite{DeRujula1, DeRujula2} has pointed out that the small value of the charge radius reported in \cite{Pohl} based on the assumption that the electric from factor of the proton $G_{E}(Q^2)$ is the dipole form. Hence the original QED formula,
\begin{equation}
L^{th}(meV)=209.9779-5.2262\langle r_{p}^2\rangle+0.00913\langle r^3_{p}\rangle_{(2)},
\label{1}
\end{equation}
is reduced into
\begin{equation}
L^{th}(meV)=209.9779-5.2262\langle r_{p}^2\rangle+0.0347\langle r^2_{p}\rangle^{3/2},
\end{equation}
because
\begin{equation}
[\langle r_{p}^{3}\rangle_{(2)}]^{2}=\frac{3675}{256}[\langle r_{p}^2\rangle]^{3},
\label{dipole}
\end{equation}
when $G_{E}(Q^2)$ is the dipole form.
(Note that in the above equations, the units of $\langle r_{p}^2\rangle$ and  $\langle r_{p}^3\rangle_{(2)}$ are fm$^2$ and fm$^3$, respectively.)
Since the experimental result is $L^{exp}$=$206.2949\pm 0.0032$~meV, accordingly they concluded that the value of the charge radius is $0.84184$ fm ~\cite{Pohl}.
However, De R\'{u}jula has argued that there is no reason to believe $G_{E}(Q^2)$ to be the dipole form. Instead he suggested that the proton
may own a large third Zemach moment about  $36.59 $ fm$^3$, which is about fifteen times larger than the value from Eq.(\ref{dipole}).
If so the value of the  charge radius extracted from Eq.(\ref{1}) will agree with the CODATA value well. Furthermore he has employed some toy model of the $G_{E}(Q^2)$ to obtain $\langle r^3_{p}\rangle_{(2)}$=36.59 fm$^3$. By this way one is able to resolve the proton size puzzle \cite{DeRujula1, DeRujula2}.

The other attempts to resolve this puzzle, for example, to recalculate the polarizability contribution \cite{Carlson} or to estimate the non-perturbative effect \cite{Thomas}, and to test the possibility of the existence of the new particle between the proton and the muon \cite{Chiang}, have been not very successful so far. The new corrections they have found are usually too small
(The only exception is so-called off-mass-shell effect advocated by \cite{Miller}). Therefore the simple solution suggested by
De R\'{u}jula seems to be worthy of further investigation.

However, the proposal of De R\'{u}jula has been severely criticized by several groups \cite{Cloet, Walcher, Ron}. First, the toy model used by De R\'{u}jula has been indeed ruled out by the recent experimental data. Furthermore,
those groups have argued that such a large value of the third Zemach moment cannot accommodate the current data of $ep$ elastic scattering. What they did, instead, is to adopt several widely used parametrizations of  $G_{E}(Q^2)$ to calculate the correspondent third Zemach moment $\langle r^3_{p}\rangle_{(2)}$ and presented them to be far smaller values than the one obtained by De R\'{u}jula.
At first glance, this objection looks very convincing.
However, as De R\'{u}jula already pointed out, the value of $\langle r^{3}_{p}\rangle_{(2)}$ is extremely sensitive to the behaviour of $G_{E}(Q^2)$ in very low $Q^2$ regime. Because there is no data between $Q^2=0$ to $Q^2=Q_{min}^2$. Therefore he argued the slim possibility of large third Zemach moment may not completely excluded yet \cite{DeRujula2}.

But one can provide a counterargument as follows: the extrapolation of $G_{E}(Q^2)$ between $Q^2$=0 and $Q^2$=$Q_{min}^2$ should be very reliable.
Because the value of $G_{E}$ at $Q^2=0$ must be one due to the fact that the electric charge of the proton is $+e$,
and its derivative $\frac{dG_{E}(Q^2)}{dQ^2}$ at $Q^2$=0 is also severely constrained by the CODATA value of $\langle r^2_{p}\rangle$.
Thus the extrapolation of $G_{E}(Q^2)$ from $Q^2$=$Q_{min}^2$ to $Q^2$=0 is supposed to be adequate enough to determine the value of $\langle r^3\rangle_{(2)}$. However this counterargument has one loophole.
If there appears a "thorn" or "dip" in $G_{E}(Q^2)$ between $Q^2$=0 and $Q^2$=$Q^2_{min}$ then the parametrizations previously used \cite{Alberico, Kelly} will no longer be able to produce an accurate value of $\langle r^{3}_{p}\rangle_{(2)}$.
Naturally one should ask that whether there exists a $G_{E}(Q^2)$ with ``thorn" or ``lump" which can generate a large $\langle r^3_{p}\rangle_{(2)}$, and at the same time, accommodate the existing $ep$ scattering data. In particular, recently a measurement of the cross section of the elastic $ep$ scattering has been carried out at Mainz, ranged from $Q^2=0.004$ GeV$^2$ to $1$ GeV$^2$ with the statistical errors below $0.2\%$ \cite{A1}. Their data has shown no sign of any ``bump". It makes any attempt to obtain large $\langle r^3\rangle_{(2)}$ by adding ``thorn" at $G_{E}(Q^2)$ to be very difficult. But in this article we will explicitly show that such a task is indeed difficult but not totally impossible.

The outline of this article is as follows. We first review the relationship between the third Zemach moment and the electric form factor $G_{E}(Q^2)$. Next we combine the ``thorn'' and some parametrizations of $G_{E}(Q^2)$ to calculate the third Zemach moment and learn
the relation between the height, width and peak position of the ``thorn" and the third Zemach moment of the proton.
Then we explicitly show that one can combine our ansatz of ``thorn" with the inverse-polynomial fit used in \cite{A1} to obtain a large third Zemach moment as De R\'{u}jula has suggested. At the same time the combined ansatz deviates from the original
inverse-polynomial fit less than $0.2\%$. Finally we present our conclusions and outlooks.

\section{Zemach moment and the electric form factors}
The conventional proton charge density is defined as the Fourier transform of the electric form factor $G_{E}(Q^2)$ in the Breit frame,
\begin{equation}
\rho_{p}(r)=\int\frac{d^3{q}}{(2\pi)^3}e^{-i{\bf q}\cdot{\bf r}} G_{E}({\bf q}).
\end{equation}
Here $Q^2=-q^2=|\vec{q}~|^2-q_{0}^2$ which is equal to $|\vec{q}~|^2$ in the Breit frame. We use the notation ${\bf q}= |\vec{q}~|$.
The following quantities are defined as
\begin{equation}
\langle r^n_{p} \rangle =\int d^{3}r r^n \rho_{p}(r).
\end{equation}
From this definition one can easily deduce that $G_{E}(0)=1$ because the $0$-th moment $\langle r^{0}_{p}\rangle$=$1$ and
$\frac{dG_{E}({\bf q}^2)}{d{\bf q}^2}|_{{\bf q}^2=0}=-\frac{1}{6}\langle r^2_{p} \rangle$. On the other hand the $n$-th Zemach moment is defined as
\begin{equation}
\langle r^{n}\rangle_{(2)}=\int d^{3}r r^n \rho_{2}(r),
\end{equation}
where $\rho_{2}(r)$ is defined as
\begin{equation}
\rho_{2}(r)=\int d^{3}r'\rho_{p}({\bf r'})\rho_{p}({\bf r'}-{\bf r})
=\int\frac{d^{3}q}{(2\pi)^3}e^{-i{\bf q}\cdot{\bf r}}G_{E}^{2}({\bf q}).
\end{equation}
After some algebra one obtains the following result \cite{Friar},
\begin{equation}
\langle r^3_{p}\rangle_{(2)}=\frac{48}{\pi}\int^{\infty}_{0}\frac{d{\bf q}}{{\bf q}^4}\left[G_{E}^2({\bf q})-\frac{{\bf q}^2}{3}
\langle r^2_{p}\rangle-1\right].
\label{Zemach}
\end{equation}
It is obvious that the third Zemach moment of the proton is dominated by the $G_{E}(Q^2)$ at very low $Q^2$.
The crucial issue here is whether there exists one form of $G_{E}(Q^2)$ which is able to produce large $\langle r^3_{p}\rangle_{(2)}$ and at the same time accommodate the current data of $ep$ elastic scattering.

\section{Relation between the Thorn in $G_{E}$ and the third Zemach moment}
In this section we assume that there is some ``thorn" or "lump" appearing in the $G_{E}(Q^2)$ in the very low $Q^2$ regime. We expect such a pathological structure to produce a large third Zemach moment.
Here we express the electric form factor as follows,
\begin{equation}
G_{E}(Q^2)=G_{E}^{(R)}(Q^2)+\Delta G_{E}(Q^2),
\end{equation}
where $G_{E}^{(R)}(Q^2)$ is some parametrization from the global fitting of the $ep$ scattering data.
On the other hand $\Delta G_{E}(Q^2)$
denotes the "thorn" on the electric form factor.
Naively, one may think that it is easier to simply add a triangle function with the height $H$ and the width $W$, whose peak is located at
$Q_{peak}^2$. However such a choice will cause a serious problem. One can calculate the associated charge density $\Delta\rho(r)$ by making the Fourier transform of $\Delta G_{E}(Q^2)$, then calculating its contribution to $\langle r^2_{p}\rangle$. However, if the triangle function is chosen then its
corresponding $\Delta \langle r^2_{p}\rangle$ calculated by
\begin{equation}
\Delta  \langle r^2_{p}\rangle=\int d^{3}r \Delta \rho(r) r^2,
\label{DR2}
\end{equation}
is actually divergent! It is due to the fact of the corresponding $\Delta \rho(r)$ actually converges slower than $1/r^4$. Hence one has to make judicious choice of the "thorn" function so that $\langle r^2_{p}\rangle$ can be kept finite. On the other hand, here we still want to employ the widely used parametrizations of $G_{E}(Q^2)$ whose value at the $Q^2=0$ have been fixed. As a result $\Delta G_{E}(Q^2=0)$ and $\frac{d\Delta G_{E}}{dQ^2}(Q^2=0)$
both have to be negligible. Moreover the influence of $\Delta G_{E}(Q^2)$ has to able to be ignored  when $Q^2\ge Q^2_{min}$. One needs figure out some
function form satisfying the above criteria. Here we present our choice as follows,
\begin{equation}
\Delta G_{E}(Q^2)=K_{1}\exp\left[-\frac{(Q^2-K_{2})^2}{K_{3}^4}\right].
\label{thorn}
\end{equation}
Here $K_{1}$ is dimensionless and the unit for $K_{2}$ and $K_{3}$ is GeV.
$K_{1}$, $K_{2}$ and $K_{3}$ denote the height, the position of the peak and the width, respectively.

To explain the result of muonic hydrogen Lamb shift one needs show that the following quantity
\begin{equation}
\Delta L(meV)=L^{theory}-L^{exp}=209.9779-5.2262\langle r_{p}^2\rangle+0.00913\langle r^3_{p}\rangle_{(2)}-206.2949,
\end{equation}
to be smaller than the experimental uncertainty $3\times 10^{-3}$ meV. If we choose the parametrizations of \cite{Alberico} or \cite{Kelly} as our $G_{E}^{(R)}$, it is easy to pick up several parameter sets of $K_{1,2,3}$ to satisfy all criteria. We list our parameter sets and their corresponding
values of $\langle r^{2}\rangle$ and $\langle r^{3}\rangle_{(2)}$ in Table (I).
One may wonder the values of $\langle r^2_{p} \rangle$ is somehow too small compared with the CODATA value: $\langle r^2_{p} \rangle$(CODATA)=0.753 fm$^2$. The reason for it is because the parametrizations we used are the results of global fitting and their value of $\langle r^2_{p} \rangle$
are somehow small. For example, $\langle r^2_{p} \rangle$(Albrico)=0.750 fm$^2$ and $\langle r^2_{p} \rangle$(Kelly)=0.744 fm$^2$.
\begin{table}[htbp]
\begin{tabular}
{|c|c|c|c|c|c|c|c|}
\hline & $G^{(R)}_{E}$ & $K_{1}$ & $K_{2}$ (GeV) &
$K_{3}$ (GeV) & $\Delta L$(meV) & $\langle r^{2}_{p}\rangle$ (fm$^2)$ & $\langle r^{3}_{p}\rangle_{(2)}$ (fm$^3)$ \\
\hline I & Alberico &0.119185 &0.08&0.0447214 &1.7$\times 10^{-4}$ &0.745137& 23.138 \\
\hline II &Alberico &0.0139929&0.08&0.053183 &1.5$\times 10^{-4}$ &0.717153 &7.11973\\
\hline III &Alberico &0.283648&0.10 &0.053183 &-2.85$\times 10^{-4}$ &0.747693& 24.5962\\
\hline IV &Alberico & 0.139056 &0.10 &0.0588566 &8.56$\times 10^{-6}$ &0.735336 &17.5272\\
\hline V &Alberico &0.720091 &0.12 &0.053183 & 1.79$\times 10^{-4}$ &0.748308 &24.9536\\
\hline VI & Kelly &0.130982 & 0.08 & 0.0422949 &1.06$\times 10^{-6}$ & 0.742016 & 21.3496 \\
\hline VII& Kelly & 0.101611 &0.08 &0.0447214 &-9.31$\times 10^{-6}$ & 0.739647&19.9926\\
\hline VIII & Kelly &0.243554 &0.10 &0.053183 &1.06$\times 10^{-4}$  & 0.741824&21.2389\\
\hline IX & Kelly &0.118405&0.10&0.0588566&-8.5$\times 10^{-4}$ &0.731308 &15.2113 \\
\hline X& Kelly & 0.627337 &0.12 & 0.053183 & 4.6$\times 10^{-6}$ &0.742354 &21.5437 \\
\hline
\end{tabular}
\caption{Our chosen parameter sets and the values of their corresponding $\langle r^{2}_{p}\rangle$ and $\langle r^{3}_{p}\rangle_{(2)}$. } \label{tab1}
\end{table}

Unfortunately the above results have all been excluded by the recent Mainz low $Q^2$ data\cite{A1}.  Nevertheless,
we have observed several important facts from the Fig (1). First, if we make the position of peak, $K_{2}$, more close to the $Q^2$=0, the height of the peak will be smaller with the same width. However, if $K_{2}$ becomes too small, it will produce relatively large $\Delta\langle r^{2}\rangle$, which is negative, thus the resultant $\langle r^{2}\rangle$ is much smaller than the CODATA value. The second important fact is as follows. With the same  $K_{2}$, the height $K_{1}$ decreases as the width $K_{3}$ increases. i.e.,  when the peak is less sharp and the height becomes smaller. Thirdly, we also find that the result is not very sensitive to the choice of the $G^{(R)}_{E}(Q^2)$ parametrizations as shown by the Table (I). These facts will instruct us to construct more realistic ansatz of $G_{E}(Q^2)$ as shown in the next section.


\begin{figure}[htbp]
\centerline{\epsfxsize 2.0truein\epsfbox{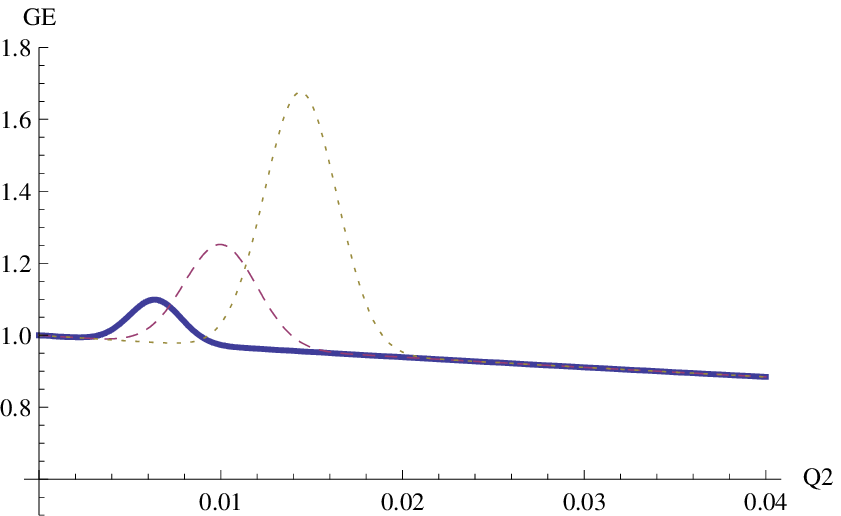} \epsfxsize
2.0 truein\epsfbox{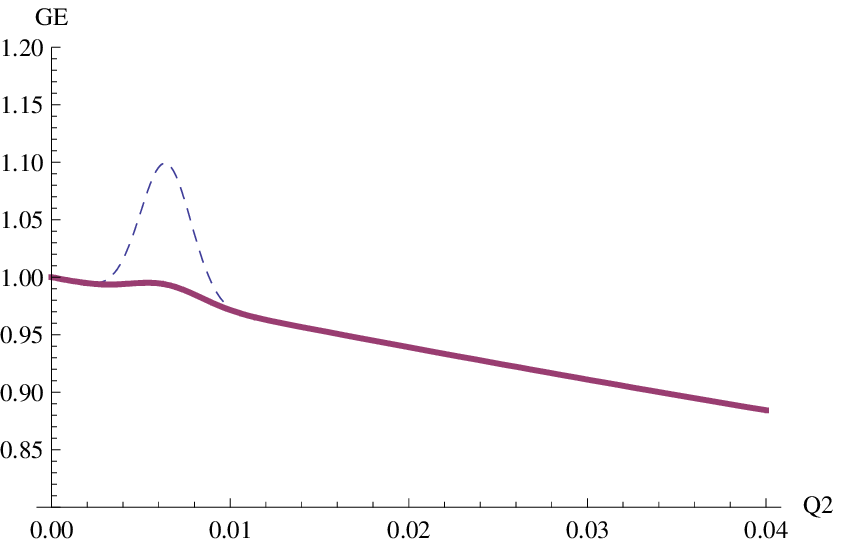}\epsfxsize 2.0 truein\epsfbox{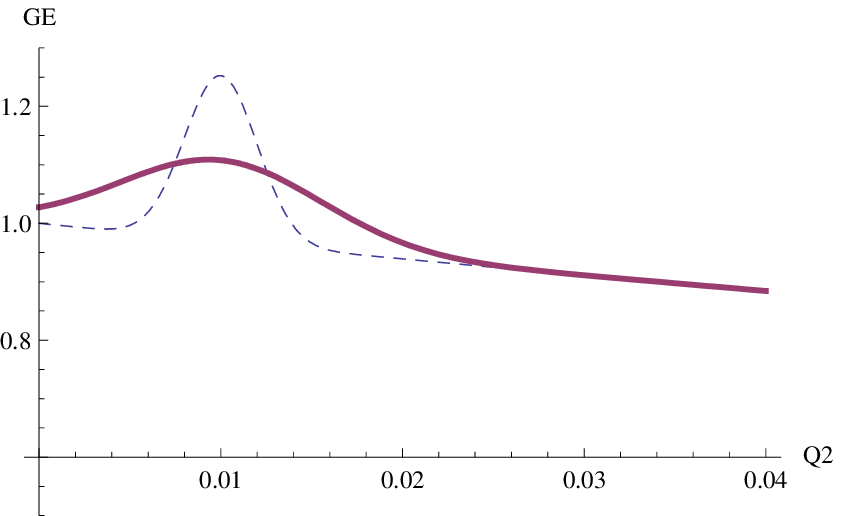}} \caption{The unit in $x$ axes is GeV$^2$. (Left)The $G_{E}(Q^2)$ in (II)(solid line), (III)(dashed line) and (V)(dotted line). Three curves correspond to the same $K_{3}$=0.053 GeV but their values of $K_{2}$ are 0.08 GeV, 0.10 GeV and 0.12 GeV respectively.
(Middle)The $G_{E}(Q^2)$ in (I) (dashed line), (II)(solid line) with
the common value of $K_{2}$=0.08 GeV. But their corresponding values of $K_{3}$ are 0.0447 GeV and 0.0532 GeV, respectively.
(right) The $G_{E}(Q^2)$ in (III)(dashed line), (IV) (solid line)
with the common value of $K_{2}$=0.10 GeV. But their corresponding values of $K_{3}$ are 0.0532 GeV and 0.0589 GeV, respectively.  }
\end{figure}

\section{Inverse-polynomial fit with a ``thorn" of $G_{E}$}
We have learned how to increase the third Zemach moment by adding the ``thorn'' in the previous section. Here we need construct a parametrization
to accommodate the recent Mainz low $Q^2$ data with $Q_{min}^2$=0.004 GeV$^2$ with the uncertainty below $0.2\%$.
In this section we show that one is able to combine our ansatz of ``thorn" with the inverse-polynomial fit used in \cite{A1,Th} to make the difference between $L^{th}$ and $L^{exp}$ to be smaller than $3\times 10^{-3}$~meV. At the same time the combined ansatz deviates from the original inverse-polynomial fit less than $0.2\%$.
The inverse-polynomial fit has been used in \cite{A1,Th}. Its explicit form is given as
\begin{equation}
G_{E}^{inv-poly}(Q^2)=\frac{1}{1+\sum_{i=1}^{7} a_{i}Q^{2i}}.
\end{equation}
with
\begin{eqnarray}
a_{1}&=&3.3615,\,a_{2}=-3.0343,\,a_{3}=29.6677,\,a_{4}=-85.6169,\, \nonumber \\
a_{5}&=&130.7053,\,a_{6}=-101.5145,\,a_{7}=34.2926.
\end{eqnarray}
Note that $G_{E}^{inv-poly}(Q^2)$ generates $\langle r^3\rangle_{(2)}$=2.96667 fm$^3$.
To combine this fit with the ansatz in Eq.(\ref{thorn}), one needs to guarantee that $G_{E}(0)=1$. Hence we modify $G^{inv-poly}_{E}$
into the following one,
\begin{equation}
G_{E}^{mod}(Q^2)=(1-\Delta G_{E}(0))G^{inv-poly}_{E}(Q^2)+\Delta G_{E}(Q^2).
\label{mod}
\end{equation}
Employing the ansatz in Eq.(\ref{mod}), one can make $\Delta L$ to be smaller than $3\times 10^{-3}$ ~meV with
the chosen parameters listed in the Table (II). The value of $\langle r^{2}_{p}\rangle$ is a little larger than CODATA value but still
be reasonable. To accommodate the very precise Mainz low $Q^2$ data, it is necessary to make $K_2$ very small. The width $K_3$ is about only half of the value used in the previous section. The height $K_{1}$ is only few percents of the ones in Table (I). However such a ``lump" generates a very large $\langle r^{3}_{p}\rangle_{(2)}$. It shows that the third Zemach moment is extremely sensitive to the detail of the electric form factor $G_{E}(Q^2)$ at very low $Q^2$ regime.

\begin{table}[htbp]
\begin{tabular}
{|c|c|c|c|c|c|}
\hline  $K_{1}$ & $K_{2}$ (GeV) &
$K_{3}$ (GeV) & $\Delta L$(meV) & $\langle r^{2}_{p}\rangle$ (fm$^2)$ & $\langle r^{3}_{p}\rangle_{(2)}$ (fm$^3)$ \\
\hline -0.0016962 &0.001&0.0221336 &5.95$\times 10^{-4}$ &0.787349 & 47.3646 \\
\hline
\end{tabular}
\caption{Our chosen parameters and the values of $\langle r^{2}_{p}\rangle$ and $\langle r^{3}_{p}\rangle_{(2)}$. } \label{tab2}
\end{table}

Moreover we define the following quantity to characterize the difference of our modified fit and the original inverse-polynomial fit.
\begin{equation}
R(Q^2)=\left[\frac{G^{mod}_{E}(Q^2)-G^{inv-poly}_{E}(Q^2)}{G^{inv-poly}_{E}(Q^2)}\right].
\end{equation}

\begin{figure}[htbp]
\centerline{\epsfxsize 3.0truein\epsfbox{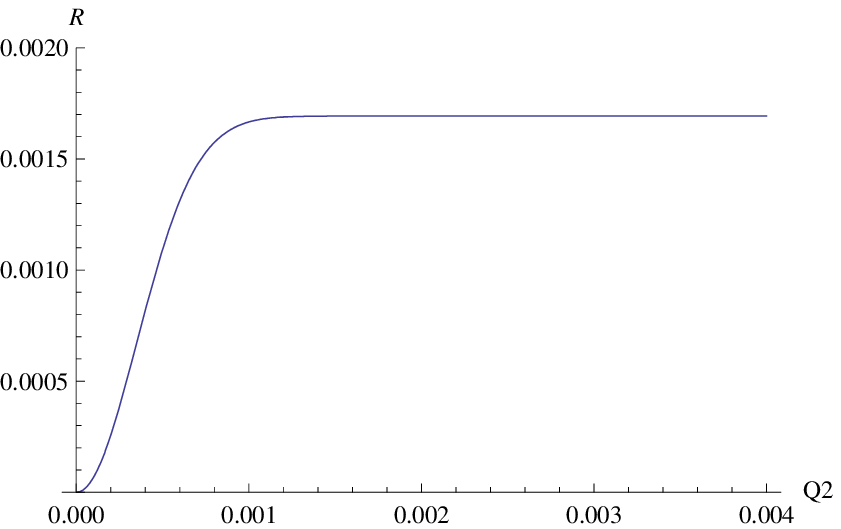}\epsfxsize 3.0truein\epsfbox{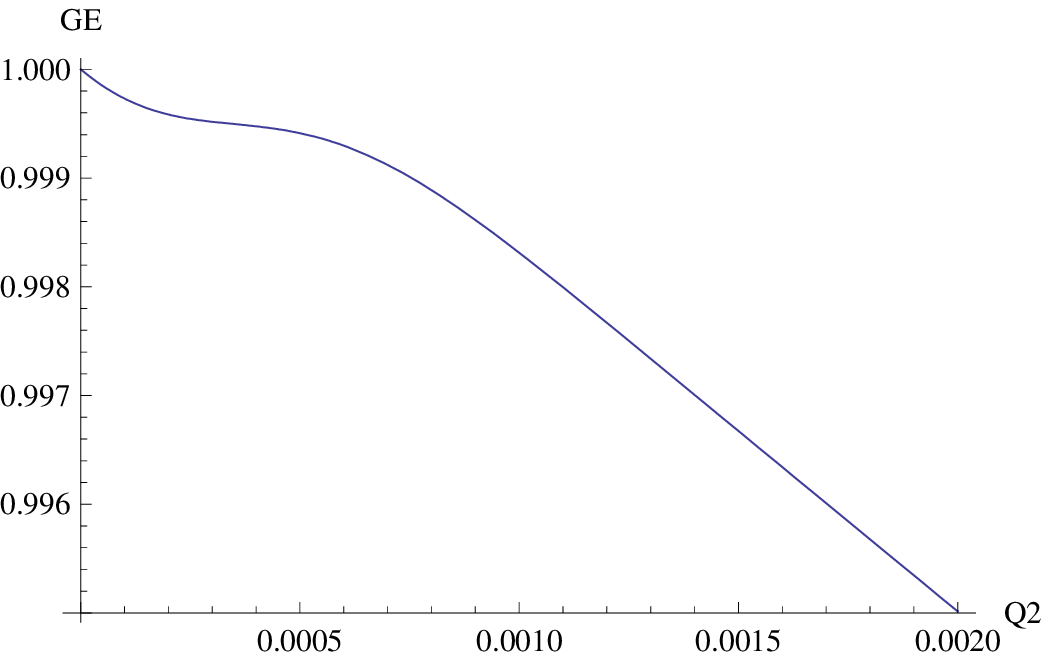}}
\label{new}
\caption{(left) The value of $R(Q^2)$ defined in the text. (right) The curve of $G^{mod}_{E}(Q^2)$.}
\end{figure}

The left panel of Fig.(2) shows that the value of $R(Q^2)$ is ranged from $0$ to $0.17\%$. It is due to the fact $\Delta G_{E}(Q^2=0)=-1.6\times 10^{-3}$. When $Q^2$ is large enough, $\Delta G_{E}\sim 0$ and $G^{mod}(Q^2) \sim (1-\Delta G_{E}(0)) G^{inv-poly}(Q^2)$. One can also observe the curve of $G^{mod}(Q^2)$. It is a very smooth ``lump" hidden at the extreme low $Q^2$ regime. The shape of this ``lump'' is depicted in the right panel of Fig.(2). It is very smooth as one can observe from the plot.

One may frown on our result here and argue that our result cannot accommodate the Mainz data satisfactorily. Indeed the ansatz in Eq.(\ref{mod}) is
quite a simple way to add a ``lump" to the existing fit. However, we emphasize that even with such a simple ansatz one can still embed a ``lump" at $G_{E}(Q^2)$ and produce a large $\langle r^{3}_{p}\rangle_{(2)}$. We believe it to be promising to improve the result with better agreement with the data by employing more sophisticated ansatz instead of the one we used here. We leave it for our future publication \cite{Kao}.

\section{Conclusion and Outlook}
In this article, we show that the third Zemach moment becomes large if there is a "thorn" or ``lump" at very low $Q^2$ regime.
Furthermore, our study show that it is possible to construct a parametrization of $G_{E}(Q^2)$ which can accommodate the existent $ep$ elastic scattering data and, at the same time,  generate a large $\langle r^3_{p}\rangle_{(2)}$ to explain the Lamb shift of the muonic hydrogen.

In this work we limit ourselves to combine the existent parametrizations of $G_{E}^{inv-poly}(Q^2)$ and a simple ansatz denoting the "thorn" by a very simple way in Eq.(\ref{mod}).
In principle one should use the following ansatz,
\begin{equation}
G_{E}(Q^2)=\frac{a_{1}+a_{2} Q^2}{1+a_{3} Q^2+a_{4} Q^4+a_{6}Q^6}+b_{1}\exp\left[\frac{-(Q^2-b_{2}^2)^2}{b_{3}^{4}}\right],
\end{equation}
to fit the $ep$ scattering data globally. There are two relations between those parameters,
\begin{equation}
1=a_{1}+b_{1}\exp\left[\frac{-b_{2}^4}{b_{3}^4}\right],
\label{c1}
\end{equation}
\begin{equation}
\frac{\langle r^2_{p}\rangle(CODATA)}{6}=-a_{2}+a_{3}+\frac{2b_{1}b_{2}^2}{b_{3}^4}.
\label{c2}
\end{equation}
Here we have four new parameters $a_{1}$ and $b_{1-3}$ with several constrains such as Eq. (\ref{c1}) and Eq.(\ref{c2}). The third constraint is
the resultant result of Eq.(\ref{Zemach}) has to be around $36$ fm$^3$.
Using the above parametrization, it is likely that one can pick up a suitable parameter set to accommodate the existent $ep$ elastic scattering data. The result by default can explain the Lamb shift of both electronic and muonic hydrogen.
We leave this task for our future publication \cite{Kao}.

Although phenomenologically a large third Zemach moment is possible,
nevertheless, there are still many challenges from theory side. For example, the very low $Q^2$ behaviour of
$G_{E}(Q^2)$ is supposed to be dominated by the chiral physics. Namely the pion cloud plays the crucial roles in the low energy regime and
one can apply Chiral Perturbation Theory ($\chi$ PT)
to calculate the electric form factor there\cite{Pineda}. The $\chi$ PT result of $\langle r^3_{p}\rangle_{(2)}$ is about $2-3$ fm$^3$, which is much smaller than ours.
We also notice the most recent estimate made by \cite{Carroll}, their conclusion disagrees with us. The reason is they insist to adopt the smooth $\rho(r)$ which guarantees $\langle r^n\rangle$ to be always finite. We only require the convergence of Eq.(\ref{Zemach}) and Eq.(\ref{DR2}) only.
These issues all remain open and need further studies.

\acknowledgements
We are very grateful to Thomas Walcher for bringing the recent Mainz low $Q^2$ data to our attention.
This work is supported by the National Science Council of
Taiwan under grants nos. NSC099-2112-M033-004-MY3 (C.W.K.).
C.W. K also acknowledges the support from the North branch of NCTS, Taiwan.

\end{document}